\documentclass[fleqn,usenatbib]{mnras}
\usepackage{graphicx}
\usepackage{lipsum}
\usepackage{amsmath}
\usepackage{breqn}

\usepackage{aas_macros}
\usepackage{amssymb}

\usepackage{color}

\title{The clustering of $z>7$ galaxies: Predictions from the \texttt{BLUETIDES} simulation}

\author[Bhowmick et al.]{
Aklant K. Bhowmick$^{1}$,
Tiziana Di Matteo$^{1}$,
Yu Feng$^{2}$,
Francois Lanusse$^{1}$
\\
$^{1}$McWilliams Center for Cosmology, Dept. of Physics, Carnegie Mellon University,
Pittsburgh PA 15213, USA\\
$^{2}$Berkeley Center for Cosmological Physics, University of California at
Berkeley, Berkeley CA 94720, USA\\
}

\date{Accepted XXX. Received YYY; in original form ZZZ}

\begin{document}

\maketitle

\begin{abstract}
  We study the clustering of the highest-$z$ galaxies (from$~\sim 0.1$
  to a few tens Mpc scales) using the \texttt{BLUETIDES} simulation
  and compare it to current observational constraints from
  \textit{Hubble} legacy and Hyper Suprime Cam (HSC) fields (at $z =
  6-7.2$).  With a box length of 400 Mpc/$h$ on each side and 0.7
  trillion particles, \texttt{BLUETIDES} is the largest volume high
  resolution cosmological hydrodynamic simulation to date ideally
  suited for studies of high$-z$ galaxies. 
  We find that galaxies with magnitude $m_{UV} < 27.7$ have a
    bias ($b_g$) of $8.1 \pm 1.2$ at $z=8$, and typical halo masses
    $M_H\gtrsim 6\times 10^{10}$ $M_\odot$. Given the redshift
    evolution between $z=8$ to $z=10$ ($b_g\propto(1+z)^{1.6}$), our inferred values of the bias and halo masses are
    consistent with measured angular clustering at $z \sim 6.8$
    from these brighter samples. The bias of fainter galaxies (in the Hubble legacy field at $H_{160} \lesssim29.5$) is
    $5.9\pm0.9$ at $z=8$ corresponding to halo masses $M_H \gtrsim
    10^{10} M_{\odot}$. We investigate directly the 1-halo term in
  the clustering and show that it dominates on scales $r \lesssim 0.1$
  Mpc/$h$ ($\Theta \lesssim 3''$) with non-linear effect at transition
  scales between the 1-halo and 2-halo term affecting scales 0.1
  $\lesssim r \lesssim $ 20 Mpc/$h$ ($3''\lesssim \Theta \lesssim
  90''$). Current clustering measurements probe down to the scales in
  the transition between 1-halo to 2-halo regime where non-linear
  effects are important.  The amplitude of the 1-halo term implies
  that occupation numbers for satellites in \texttt{BLUETIDES} are
  somewhat higher than standard HODs adopted in these analyses (which
  predict amplitudes in the 1-halo regime suppressed by a factor
  2-3). That possibly implies a higher number of galaxies detected by
  JWST (at small scales and even fainter magnitudes) observing these
  fields.

\end{abstract}


\section{Introduction}
\label{introduction}
Advancement of observational techniques and the emergence of more
powerful telescopes have pushed the observational frontiers to
increasingly higher redshifts. Detecting the earliest galaxies holds
the key to understanding the role of the first galaxies in triggering
the epoch of reionization. The study of Lyman Break Galaxies (LBGs) is
one such probe with a number of candidate objects detected by the HST
fields to redshifts as far as $z=11$
\citep{2016ApJ...819..129O,2016ApJ...830...67B,2016MNRAS.459.3812M}. Upcoming
satellites such as the \textit{James Web Space Telescope} (JWST) and
the \textit{Wide Field Infrared Survey Telescope} (WFIRST-AFTA) are
expected to detect a large number of objects at redshifts 8-15
\citep{2016MNRAS.463.3520W}. \par
Performing clustering analysis for these distant galaxies ($z\geq8$)
poses a major challenge due to 1) not having enough objects to obtain
reliable statistics 2) presence of cosmic variance for narrow fields
of survey \citep{2008ApJ...676..767T,2014ApJ...796L..27R,
  2004ApJ...600L.171S}. For instance, HST WFC 3 has a 2 arcmin IR
field of view, and is therefore subjected to substantial amount of
cosmic variance. JWST will observe at greater depths and hence detect
more objects, but will likely have a similar degree of cosmic
variance. WFIRST will have a survey area of $2200 \ \textrm{deg}^2$
which is comparable to that of ground based survey, and will have a
depth comparable to that of HST-Ultra Deep
Field. \cite{2013arXiv1305.5422S} predicts a limiting magnitude of $26
\ (10\sigma)$ and $26.75 \ (5\sigma)$ respectively for the
WFIRST. WFIRST observations will be amongst the first to reliably
capture the clustering of the brightest galaxies at $z\sim8,9,10$,
some of the predictions of which have been made by
\cite{2016MNRAS.463.3520W}. \par
Despite the foregoing difficulties, clustering studies have been
performed for medium to high galaxies within the later half of the
reionization epoch and after. \cite{2015MNRAS.454..205I} analyzed the
clustering properties at $z\sim2$ using the data taken at the United
Kingdom Infrared telescope and Canada-France-Hawaii Telescope
(CFHT). \cite{2009A&A...498..725H} obtained the angular correlation
functions of Lyman Break Galaxies (LBGs) at $z\sim3-5$ using the
observations of Canada-France-Hawaii Telescope Legacy Survey
(CFHTLS). \cite{2001ApJ...558L..83O} performed clustering studies at
$z\sim4-5$ using the data taken at the Subaru deep
survey. \cite{2004MNRAS.349..281M} took a compilation of a number of
surveys and studied the evolution of galaxy clustering from $z\sim0$
to $z\sim5$. Similarly, \cite{2005ApJ...619..697A} compiled star
forming galaxies from multiple surveys and studied their spatial
clustering at redshifts $1.4<z<3.5$. Owing to the data collected by
the Hubble Space Telescope Wide field camera 3 (HST-WFC3), the
observation frontier for large scale clustering has been pushed to
$z\sim7$. \cite{2014ApJ...793...17B} estimated the angular clustering
of LBG galaxies at redshifts upto $z\sim7.2$ using a combined sample
of LBGs from HST-WFC3, Hubble eXtreme Deep Field (XDF) and CANDELS
surveys; and thereafter for the first time estimates the large scale
linear bias at $z\sim7.2$ by fitting the angular correlation function
to a power-law model. \cite{2016ApJ...821..123H} and
\cite{2017arXiv170406535H} further expanded the compilation by
including the Hubble Frontier Field (HFF) and the HSC SSP(Subaru) data
and made improved estimates of the angular clustering and the large
scale bias.\par
The clustering measurements of observed galaxy samples enable
  us to simultaneously constrain cosmology, as well as physics of
  galaxy formation; this is typically done using Halo Occupation
  Distribution (HOD) modelling \citep{2002ApJ...575..587B}.
  Incorporating the HODs into the framework of the halo model
  \citep{2002PhR...372....1C}, one can study galaxy clustering and fit
  model parameters to observational measurements, thereby determining
  the average halo masses and biases of the observed galaxy subsamples
  i.e. the galaxy-dark matter halo connection. This has been
  extensively applied to low ($z=0-1$) to medium ($z=2-5$) redshift
  measurements
  \citep{2005ApJ...630....1Z,2007ApJ...667..760Z,2013MNRAS.430..725V,doi:10.1046/j.1365-8711.2002.04959.x,doi:10.1111/j.1365-2966.2004.07253.x,2013ApJ...774...28B,2007A&A...462..865H}). Recent
  studies have started focusing on the application of HOD modelling to
  galaxies within the reionization epoch
  \citep{2016ApJ...821..123H,2017arXiv170406535H,2017arXiv170203309H}.

Semi-analytical (SA) models \citep[and references therein]{2012NewA...17..175B,2015A&A...575A..33C,2017MNRAS.471.1401C,2017arXiv170802950H} have also been extensively used to predict the clustering
  as well as other intrinsic properties of galaxies.  Several studies
  using SA models have been performed to predict galaxy clustering at
  high redshifts.  \cite{1998ApJ...498..504B} constructs mock galaxy
  catalogs $z\sim3$ and reports broad agreement with observations
  existing at that time. \cite{2016MNRAS.461..176P} studies clustering
  of LBGs at $z\sim4$ using semi analytical models and finds good
  agreement with the measurements of Hubble eXtreme Depp Field (XDF)
  and CANDELS fields. \cite{2006ApJ...637..631K} also analyzes
  clustering of LBG galaxies from the Subaru deep field at $z\sim4,5$
  and achieves good agreement with the results obtained from a mock
  galaxy catalog constructed using semi-analytical
  modelling. \cite{2013MNRAS.435..368J} builds SA models to predict
  clustering of high redshift galaxies and achieves excellent
  agreement with observations from $z=3-6$. \cite{2016MNRAS.463..270J}, \cite{2017MNRAS.469.4428J} use N-body simulations to study clustering of dark matter halos at high redshifts ($z\sim3-5$) and reveal the existence of non-linear bias in the two-halo regime at these redshifts. More recently, \cite{2017MNRAS.472.1995P} used N-body + semianalytic models to make predictions of clustering of LBGs upto $z\sim8$, and shows that they agree well with current measurements upto $z\sim7$. As we shall show later, their predictions at $z\sim8$ agree well with \texttt{BLUETIDES}. 

Recently there has been tremendous progress in cosmological
hydrodynamic simulations of galaxy formation. Capturing (as much of)
the relevant physics at small scales while simulating statistical
populations of galaxies demands a combination of having a large
volume, as well as having large enough number of resolution elements.
Several large scale simulations have been performed in the recent
years in order to study the formation of the first galaxies and their
roles in the epoch of reionization. \texttt{MASSIVE BLACK I}
\citep{2012ApJ...745L..29D} has a box length of $533 \textrm{Mpc}/h$
and was run to $z=4.75$. Its successor, \texttt{MASSIVE BLACK II} had
a smaller box length of $100 \textrm{Mpc}/h$ and therefore had a
higher resolution \citep{2015MNRAS.450.1349K}; it was run to
$z\sim0$. \texttt{MASSIVE BLACK II} illustrated that the inclusion of
baryons have strong effects on the abundances of dark matter halos,
which motivated more simulations of such kind. \texttt{ILLUSTRIS}
\citep{2015A&C....13...12N} was similar to \texttt{MASSIVE BLACK II}
in terms of simulation volume as well as resolution, but with improved
modelling for SN II feedback
\citep{2010MNRAS.406..208O}. \texttt{EAGLE} simulation
\citep{doi:10.1093/mnras/stu2058} is also comparable to that of
\texttt{ILLUSTRIS} and \texttt{MASSIVE BLACK II}, but with improved
modelling of sub-grid physics which lead to better agreement with
observations. \par

In order to study clustering at high redshifts, large volume
simulations are required (the high-z galaxy populations are
rare-bright objects). Here we use the \texttt{BLUETIDES} simulation
\citep{2016MNRAS.455.2778F} with a box side qlength of 400 Mpc/$h$ and
has a particle load 10 times that of \texttt{MASSIVE BLACK I}. Due to
high particle load, the resolution is comparable to that of
\texttt{MASSIVE BLACK II} or \texttt{ILLUSTRIS}, which makes it the
largest high resolution cosmological simulation performed to date. The
flipside to this is that the simulation is so computationally
demanding that it has currently been run only down to $z\sim8$.  While
attempts are being made to push the simulation further down to lower
redshifts here we wish to first further validate \texttt{BLUETIDES} by
examining the predictions for galaxy clustering and compare to current
constraints.
The photometric properties of the \texttt{BLUETIDES}  galaxies have been
calculated in \cite{2016MNRAS.460.3170W} using five different stellar
population synthesis models. The predicted luminosity functions and
star formation rates show good consistency with the currently
available observations at $z\sim8,9,10$ \citep{2016MNRAS.455.2778F};
the resolution is high enough to obtain the slope of the faint end of
the luminosity function which is consistent with
\cite{2014ApJ...793..115B}. \texttt{BLUETIDES} also consistently
predicted the properties of GN-z11 \citep{2016ApJ...819..129O} which
is the farthest galaxy observed to date \citep{2016MNRAS.461L..51W}.
\texttt{BLUETIDES} forecasts for the upcoming WFIRST satellite have
been made by \cite{2016MNRAS.463.3520W} and results look promising
with about $10^6$ predicted galaxies from $z\sim8$ upto $z\sim15$. In
the large volume of \texttt{BLUETIDES} it has also been possible to
study the formation of the first quasars, supermassive black holes
\citep{2017MNRAS.467.4243D}.\par
In this work, we perform a detailed analysis of the clustering of
\texttt{BLUETIDES} galaxies at redshifts $z\sim 8,9,10$. We predict clustering properties of observed galaxy samples at $z=8$ \citep[hereafter B15]{2015ApJ...803...34B}, and estimate 
their typical biases and threshold halo masses. We investigate the redshift evolution of the clustering and compare to current clustering measurements at lower redshifts \citep[hereafter H16]{2016ApJ...821..123H}. We compare our clustering
results, particularly the one-halo term, to the predictions of an HOD
model, and highlight some differences between them at various length
scales.

\section{The Bluetides simulation}
\label{simulation}
\begin{figure*}
  \includegraphics[width=17 cm,height=17 cm]{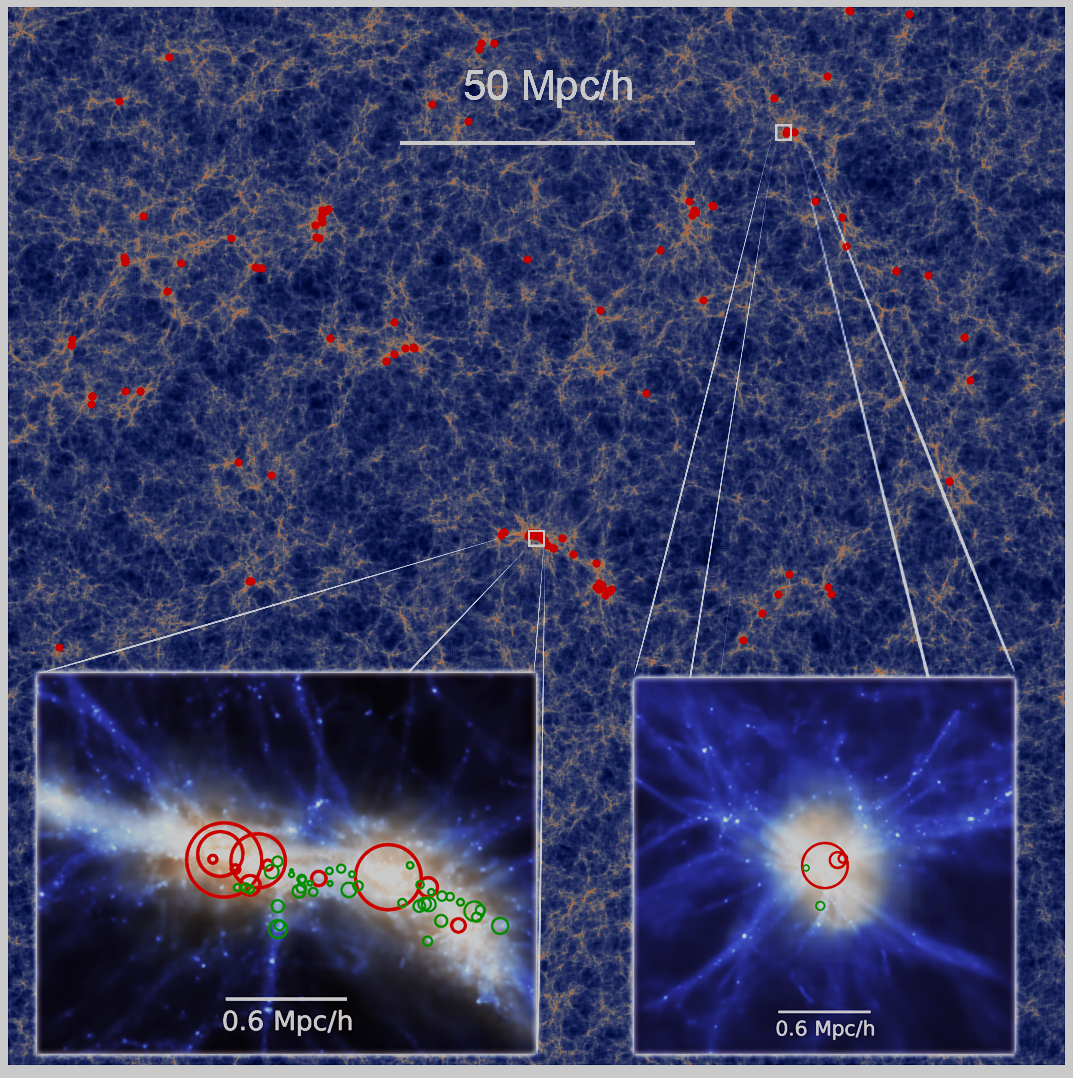}
  \caption{BACKGROUND PANEL: Large scale structure of the dark
      matter distribution at $z=8$ over a $200 \times 200\
      (\textrm{Mpc}/h)^2$ panel of thickness 2 Mpc/$h$. The dark
      (blue) and bright (yellow) regions indicate lower and higher
      dark matter densities respectively. The red dots show the
      positions of subhalos hosting galaxies of absolute magnitude
      $m_{UV}\lesssim27.7$ (threshold magnitude of bright subsamples of LBG
      galaxies from the Hubble legacy and HSC fields at $z\sim7$ for
      which H16 performs clustering measurements). INSET PANELS: The
      insets show the gas density profiles of two Friends-of-friends
      (FOF) halos of masses $3.8\times10^{12}\ M_{\odot}/h$ (left) and
      $7.4\times10^{11}\ M_{\odot}/h$ (right). Brighter colors
      represent higher gas temperature. The red circles represent
      positions and the scaled virial radii of the subhalos which were
      marked as red dots in the background panel. The green circles
      represent subhalos hosting fainter galaxies expected to be
      observed in the next generation deep field surveys of JWST
      ($M_{UV}<-16$) \citep{2011MNRAS.414..847S}.}
\label{screenshot}
\end{figure*}

\begin{table*}
\begin{tabular}{|c|c|c|}
\hline
Description&Parameter&Value\\
\hline
Dark energy density&$\Omega_{\lambda}$&0.7186\\
Matter density(Dark matter+Baryons)&$\Omega_{0}$&0.2814\\
Baryon density&$\Omega_{b}$&0.0464\\
Dimensionless Hubble parameter&h&0.697\\
Linear mass dispersion at $r=8 MPc h^{-1}$&$\sigma_8$&0.82\\
Mass of dark matter particle&$M_{\textrm{DM}}$&$1.2\times10^{7}M_{\odot}/h$\\
Mass of a gas particle&$M_{\textrm{GAS}}$&$2.36\times10^{6}M_{\odot}/h$\\
Seed mass of Black hole&$M_{\textrm{BH}}$&$5\times10^{5}M_{\odot}/h$\\
Gravitational soothening length&$\epsilon_{grav}$&$1.5Kpc h^{-1}$\\
\hline
\end{tabular}
\caption{Parameters for the \texttt{BLUETIDES} simulation. Cosmology has been taken from the results of the Wilkinson Micrwave Anisotropy Probe nine-year data release \protect\citep{2013ApJS..208...19H}}
\label{bluetides_cosmology}
\end{table*}
\texttt{BLUETIDES} was run on the Bluewaters super-computer at the
National center for Computing applications. It essentially required
the entire infrastructure of Bluewaters which comprises of 20250 nodes
(648000 core equivalents). \texttt{BLUETIDES} is built upon on the
precursor code \texttt{MP-GADGET}, to which substantial improvements
were made to the parallel infrastructure in order to run at Petascale
and exploit the full computing power of Bluewaters. Table
  \ref{bluetides_cosmology} summarizes the important parameters used
  in \texttt{BLUETIDES}; a full discussion on the specific features
is published in \cite{2016MNRAS.455.2778F}. The Bluetides simulation
employs the pressure-entropy formulation of Smooth Particle
Hydrodynamics. The simulation is performed inside a comoving
rectangular volume having a length of $400\ \textrm{Mpc}/h$ on each
side. The total number of particles is about 0.7 trillion. The ensuing
high resolution enables us to perform an extensive study of the
formation of the first galaxies which may have driven reionisation,
via detailed modelling of sub-grid physics described in
\ref{subgrid_physics}. In addition to the high resolution, the
extraordinarily high volume further allows us to study the large scale
clustering of galaxies and dark matter halos. Overdensities were
mapped using the Friend-of-Friends (FOF) groups identifier
\citep{1985ApJ...292..371D} with a linking length of 0.2 times the
average spacing between the particles. The first FOF groups are
identified at $z\approx17$ and their evolution is traced to
$z\approx8$. The simulation is about 300 times larger than the largest
observational survey at $z\approx 8$
\citep{2010ApJ...714L.202T,2011ApJ...727L..39T}. The substructure
within the primary FOF groups were identified using
\texttt{ROCKSTAR-GALAXIES} \citep{behroozi}. While there are ongoing
efforts to push the simulation further down to low redshifts ($z<8$),
the focus of our present work is on galaxies at redshifts 8,9 and
10. A visualization of the large-scale structure at $z=8$ is shown in
Figure \ref{screenshot}. The base panel is a $200\times200\
(\textrm{Mpc/h})^{2}$ slice of thickness 2 Mpc/$h$, which shows the
positions of the subhalos hosting galaxies with $M_{UV}<-19.3$ within
the large scale density profile of dark matter. We also zoom in to
reveal the gas temperature profiles of the substructures (shown as
insets) of two FOFgroups of masses $3.8\times10^{12}\ M_{\odot}/h$
(left) and $7.4\times10^{11}\ M_{\odot}/h$ (right). The red circles
show the positions and virial radii of galaxies with
$M_{UV}<-19.3$. As we shall see later, this threshold absolute
magnitude corresponds to one of the observed samples obtained from HSC
and Hubble legacy fields at $z\sim7$, and in the following sections we
are going to present a detailed clustering analysis of the entire
sample of these galaxies residing within the \texttt{BLUETIDES}
volume. We have also shown in green circles, an additional set of
subhalos hosting fainter galaxies which are expected to be observed by
the future deep field surveys of JWST. The galaxies in these FOFGroups
have stellar masses ranging from $10^7$ to $10^{10}\ M_{\odot}/h$ and
star formation rates ranging from $10^{-2}$ to $10^{3}\ M_{\odot}\
yr^{-1}$.

In order to calculate the UV_luminosities, we use the
  \texttt{PEGASE-v2} \citep{1997A&A...326..950F} stellar population
  synthesis (SPS) model along with the \cite{1538-3873-115-809-763}
  initial mass function. The cumulative spectral energy distributions
  (SEDs) for each galaxy is calculated from SEDs for each star
  particle (as a function of stellar age and metallicity). The
choice of the SPS model or the initial mass function makes a
difference of at most 0.2 dex in the predicted luminosities
\citep{2016MNRAS.460.3170W}, which is not large enough to
significantly impact our final results. For a complete discussion of
the photometric properties of \texttt{BLUETIDES} we urge the readers
to refer to \cite{2016MNRAS.460.3170W}.
\subsection{Sub grid physics modelling} 
\label{subgrid_physics}
A number of sub grid physics models were invoked in order to study
their influence on galaxy properties. To incorporate star formation, a
multi star formation model as prescribed in
\cite{doi:10.1046/j.1365-8711.2003.06206.x} has been implemented.  Gas
cooling through radiative transfer and metal cooling are implemented
using models of \cite{1996ApJS..105...19K} and
\cite{2014MNRAS.444.1518V} respectively. We account for the effects of
molecular hydrogen in the estimation of the star formation rate at low
metallicities. Formation of molecular hydrogen is modelled by the
prescription of \cite{2011ApJ...729...36K}. The modelling of type II
Supernovae (SNII) feedback is performed according to
\cite{2010MNRAS.406..208O}. The modelling of Active Galactic Nuclei
(AGN) is the same as that performed in the \texttt{MASSIVE BLACK} I
and II simulations \citep{2012ApJ...745L..29D,2015MNRAS.450.1349K}. Black holes of masses $5 \times 10^5
M_{\odot}/h$ are seeded at the local potential minima of halos with
masses exceeding $M_{H}=5 \times 10^{10} M_{\odot}/h$. The resulting
feedback energy is deposited within a spherical volume of radius twice
that of the SPH smoothing kernel of the black hole. The large volume
of the simulation also enables us to estimate the effects of "patchy"
reionization, wherein the semianalytical modelling is applied to
identify spatial regions which would be ionized at a given redshift;
for further details we urge the readers to refer
\cite{2016MNRAS.455.2778F} and \cite{0004-637X-776-2-81}.


\subsection{Clustering}
The large volume and high resolution of the simulation enables us to
study clustering of galaxies and dark matter halos at large scales as
well as small scales. The clustering may be quantified by the spatial
correlation function $\xi(r)$ (in configuration space), or the 3D
power spectrum $P(k)$ (in Fourier space). The spatial correlation
measures an excess in the pair-wise distribution of particles compared
to the random uniform distribution and may be defined as
\begin{equation}
dN= n (1+\xi(r)) dV
\end{equation}
where $dN$ is the total number of particles at a distance $r$ paired
with a particle sitting at the origin, $n$ is the volume density for a
uniform distribution and $dV$ is the volume of the spherical shell at
radius $r$. For the bluetides data, we compute the galaxy and dark
matter correlation functions ($\xi_{gg}(r)$ and $\xi_{dark}(r)$ via
brute force massively parallel pair counting using the
\texttt{KDCOUNT} package (https://github.com/rainwoodman/kdcount).\par
The linear and non-linear regimes can be identified by computing the
scale-dependent galaxy bias defined as,
\begin{equation} 
  b_g =\sqrt{\xi_{gg}/\xi_{\rm{dark}}}.
  \label{bias}
\end{equation}
At large scales we expect the bias to be constant
\citep{1996MNRAS.282..347M,1999MNRAS.308..119S,2010ApJ...724..878T}
whereas at small scales there is an onset of scale dependence in the
bias due to non-linear effects \citep{2016MNRAS.463..270J}. We also
expect a steep increase in the bias as we go down to scales smaller
than the typical halo radii due to a rise in the one halo
contribution. \par
The angular correlation function $\omega(\Theta)$ can be obtained by
projecting $\zeta(r,z)$ using the Limber projection formula,
\begin{equation}
\omega(\Theta)= 2\int N^{2}(z) \int_{u_{\rm{min}}}^{u_{\rm{max}}} \xi(r,z)
\frac{dz}{dx} du dz 
\label{limber_from_xi}
\end{equation}
where $u^2\equiv r^2-x^2\Theta^2$, $x(z)$ is the comoving distance and $N(z)$ is the
normalized redshift distribution.

\section{Halo Occupation Distribution (HOD) modelling}
\label{HOD}
The main results in this work are based on analysing the clustering of
high-z galaxies in the \texttt{BLUETIDES} simulation. We will however also
attempt to reproduce the HOD models used by H16 as well. This allows us
to compare directly simulations and HODs for these high-z galaxies.

For completeness, here we review briefly the standard assumptions used when building
HODs and in particular those used in H16.  HOD modelling is a commonly
used technique to obtain the galaxy halo connection from clustering
measurements. The key assumption is that the HOD i.e. the probability
distribution of the number of galaxies for a given halo of mass $M_H$,
is dependent only on the halo mass. The occupation $N(M_H)$ is
therefore given by
\begin{equation}
N(M_H)=N_c(M_H)+N_s(M_H)
\end{equation} 
where $N_c$ and $N_s$ are the mean number of central and satellite
galaxies respectively. The mean occupations of central and satellite
galaxies are given by,
\begin{equation}
\left<N_c(M_H)\right>=\frac{1}{2}\left[1+\textrm{erf}\left(\frac{\log{M_H}-\log{M_{\textrm{min}}}}{\sigma_{\log{M}}}\right)
\right],
\label{central}
\end{equation}

\begin{equation}
\left<N_s(M_H)\right>=\left<N_c(M_H)\right>\left(\frac{M_H-M_0}{M_1^{'}}\right)^\alpha
\label{satellite}.
\end{equation}

$N_c$ is effectively a smoothened step function with
$\sigma_{\log{M}}$ being the width of transition between 0 to 1. The
non-zero width of the transition is supposed to quantify the scatter
in the Halo mass-luminosity relation. $M_{\textrm{min}}$ is the mass
of the host halo at which 50 percent of the host halos contain a
central galaxy. $N_s$ is assumed to be power law for halos above a
cutoff mass of $M_0$; $\alpha$ is the power law exponent and $M_1^{'}$
is referred as the slope of the power law.

Assuming that both $N_c$ and $N_s$ are 
independent of each other and $N_s$ is a Poisson distribution, the first moment of $N(M_H)$ is given
by
\begin{equation}
\left<N(N-1)\right>=\left<N_c(M_H)\right>\left<N_s(M_H)\right>+\left<N_s(M_H)\right>^{2}.
\end{equation}    
One can similarly obtain higher order moments but in the current
context, we focus on the two point correlation function which is
completely determined by the zeroth and first order moments of the
HOD. \par
The mean density of galaxies at redshift $z$ is given by
\begin{equation}
n_g=\int \textrm{DC} \frac{dn}{dM_H} (M_H,z) N(M_H) dM_H
\end{equation}
where $\frac{dn}{dM_H} (M_H,z)$ is the halo mass function. The factor DC is referred
to as the Star formation \textit{duty cycle}, defined as the
probability of observing an existing star forming galaxy. DC is often
less than unity because of the episodic nature of star formation
activity \citep{2009ApJ...697.1493S}.  While DC is an important
parameter in an HOD model, it does not affect galaxy clustering as
along as it is independent of Halo mass. Thus, in our work we assume
DC=1 as is done in \cite{2017arXiv170406535H}.
\par The power spectrum is written as
\begin{equation}
P(k,z)=P^{1h}(k,z)+P^{2h}(k,z)
\end{equation}
where 1h and 2h represents the one-halo and two-halo contributions
respectively. The one-halo term is further divided into
central-satellite (cs) and satellite-satellite (ss) components,
\begin{equation}
P^{1h}(k,z)=P^{1h}_{\textrm{cs}}(k,z)+P^{1h}_{\textrm{ss}}(k,z),
\end{equation}

\begin{multline}
  P^{1h}_{\textrm{cs}}(k,z)=\frac{2}{n_g^2}\int
  \left<N_c(M_H)\right>\left<N_s(M_H)\right> \frac{dn}{dM_H} (M_H,z)\\
  u (k,M_H,z) dM_H,
  \label{eqn_1h_cs}
\end{multline}

\begin{equation}
  P^{1h}_{\textrm{ss}}(k,z)=\frac{1}{n_g^2}\int \left<N_s(M_H)\right>^{2}
\frac{dn}{dM_H}(M_H,z) u(k,M_H,z)^{2} d M_H
\label{eqn_1h_ss}
\end{equation}
 The two-halo term is given by
\begin{multline}
  P^{2h}(k,z)=P_m(k,z)\biggl[\frac{1}{n_g}\int \left<N(M_H)\right>
  \frac{dn}{dM_H}(M_H,z)\\ u(k,M_H,z) b_g(M_H,z) d M_H\biggr]^2
  \label{eqn_2h}
\end{multline}
where $P_m(k,z)$ is the dark matter power spectrum, $b_g(M_H,z)$ is
the large scale/linear bias of halos of mass $M_H$, $dn/dM_H$ is the mass function, $u(k,M_H,z)$ is the fourier transform of the satellite galaxy density profile normalized by the halo mass. We keep our considerations identical to that of H16. We use the mass function from
\cite{behroozi} which essentially applies a correction factor to
\cite{2008ApJ...688..709T} for high redshifts, $b_g(M_H,z)$ is obtained from \cite{2010ApJ...724..878T}. $u(k,M_H,z)$ presumably traces the darkmatter distribution which is assumed to be NFW \citep{1996ApJ...462..563N} with a mass-concentration relation taken from \cite{2003ApJ...590..197S}.\par
The angular correlation function $\omega(\Theta)$ 
can be obtained from $P(k)$ using the Limber projection for the power spectrum,
\begin{equation}
\omega(\theta)=\int dz N^{2}(z)\left(\frac{dx}{dz}\right)^{-1}\int dk
\frac{k}{2\pi}P(k,z)J_0(x(z) k \theta)
\end{equation}
where $J_n$ is the $n^{th}$ order Bessel function of the first kind,
$N(z)$ is the normalized redshift distribution and $x(z)$ is the
comoving distance as a function of redshift.

\section{Results}

In this section, we show some of the main properties of the galaxies
in \texttt{BLUETIDES} as a function of their halo mass. We then move
to the results on the clustering and compare to current observational
measurements.

\subsection{Galaxy-Halo connection in \texttt{BLUETIDES}  }

\label{GH_connection}
\begin{figure*}
  \includegraphics[width=\textwidth,height=7cm]{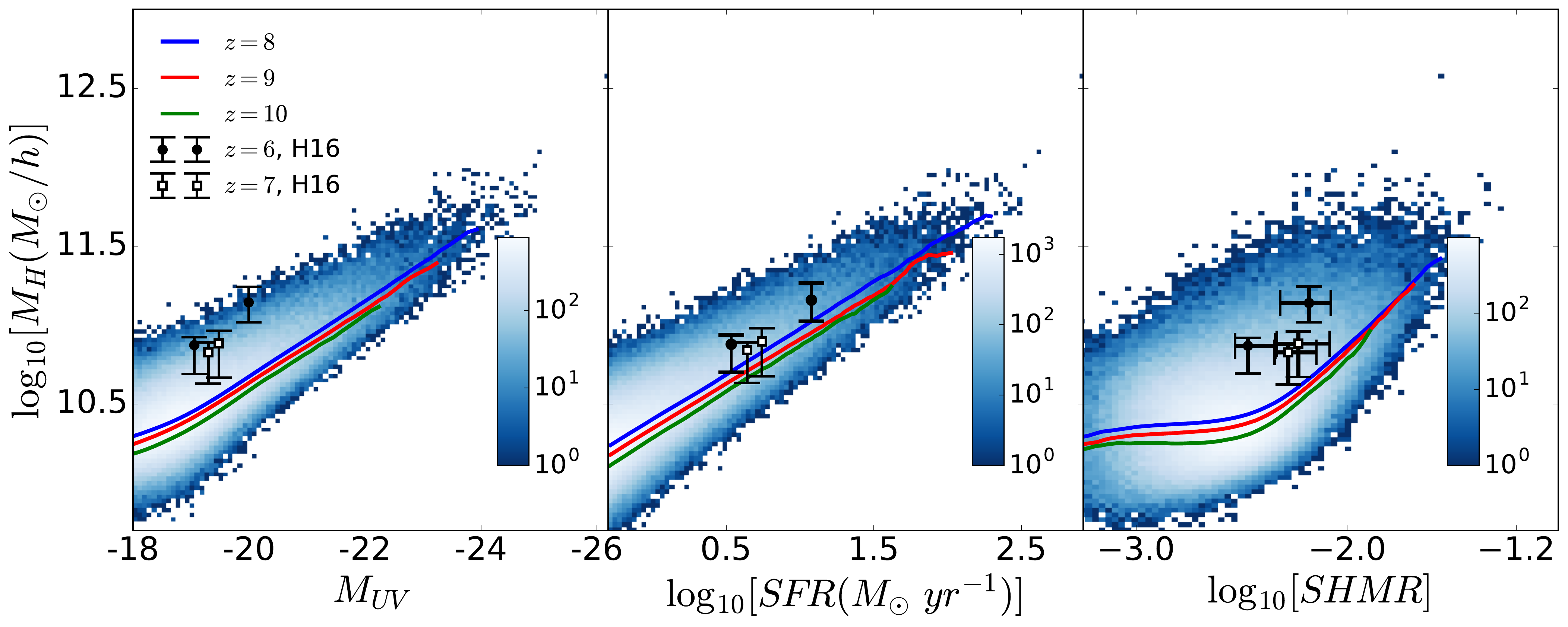}
  \caption{Relationships between the properties of the complete sample
    of galaxies from the Bluetides simulation. $M_H$ is the halo mass,
    $M_{UV}$ is the UV absolute magnitude, SFR is the Star Formation
    rate and SHMR is the Stellar to Halo Mass ratio. The 2D histograms
    are at $z=8$ with the lower and higher density regions represented
    by the darker and lighter shades of blue respectively. The solid
    lines represent the mean variations of the relationships at
    redshifts 8,9,10. The observational measurements at redshifts 6
    and 7 are published in \protect\cite{2016ApJ...821..123H} (hereafter
    referred as H16).}
\label{complete_samples}
\end{figure*}

\begin{figure*}
  \includegraphics[width=\textwidth,height=7cm]{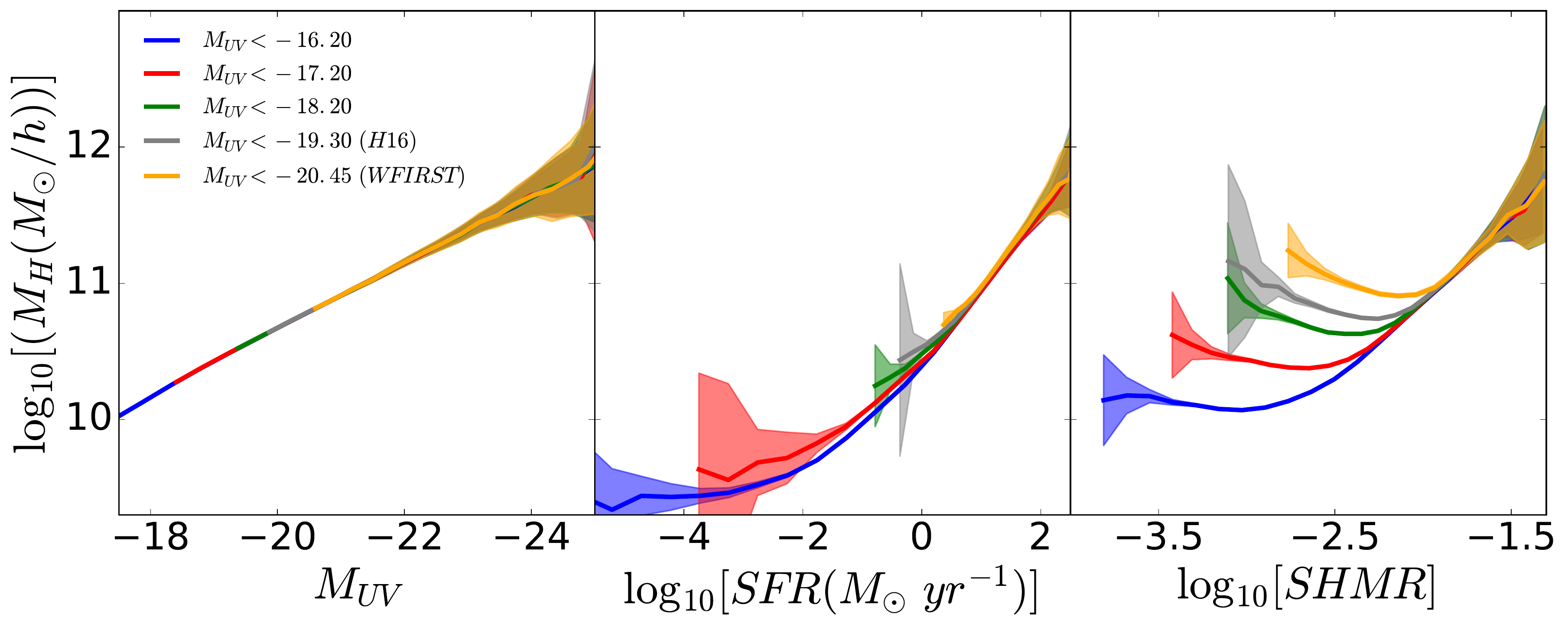}
  \caption{Variation of the relationships between various galaxy
    properties at $z=8$ for various absolute UV magnitude cuts. Solid
    lines represent the mean variation and the shaded regions
    represent the $1\sigma$ Poisson errors.}
\label{muv_selected_samples}
\end{figure*}
Figure $\ref{complete_samples}$ shows three relationships between the
the halo mass $M_{\rm H} $, UV absolute magnitude ($M_{UV}$), Star formation rate (SFR) and stellar to halo
mass ratio (SHMR) for the full galaxy population in
\texttt{BLUETIDES}. The 2D histograms show the galaxies at $z=8$ while
the solid lines are the mean relations at $z=8,9,10$ (blue, red and
green respectively). The observational data points are from H16 at $z=
6,7$. The halo masses in the observed data are inferred using HOD
modelling of the clustering. The \texttt{BLUETIDES} simulation clearly
reproduced the overall properties of the observed galaxies.  From the
$M_{\rm H}$ vs $M_{UV}$ relation, we see that at fixed $M_{UV}$ the host
halos have higher masses at lower redshifts.  $M_{\rm H}$ vs SHMR also
indicates larger halo mass for a fixed SHMR with decreasing redshifts.
This indicates that at these redshifts, the halo mass assembly
proceeds more efficiently than stellar mass.
The foregoing trends are in broad agreement with the H16
observations for $M_H=10^{11}\ M_{\odot}$ at $z=6,7$, (which continues to
be so till $z=4$ but for $z<4$, H16 measurements show a reversal in this
trend).
Updated measurements in \cite{2017arXiv170406535H} at
$M_H=10^{11}\ M_{\odot}$ show the same trends; however no evidence of
evolution was found in $M_H=10^{12}\ M_{\odot}$. This is also
consistent with \texttt{BLUETIDES} where we can see on the $M_H$
vs. SHMR plane of Figure $\ref{complete_samples}$ that the redshift
evolution gradually becomes smaller at higher halo masses. All of
these can provide significant insights into the interplay between the
relevant processes at low and high redshifts; we defer this discussion
to Section {\ref{discussion}}. \par

Since the observed galaxy samples are magnitude-limited it is
important to investigate whether the relations shown in Figure
\ref{complete_samples} change as function of magnitude limit. In
Figure \ref{muv_selected_samples}, we consider galaxy samples selected
for different absolute magnitude cuts and plot the same set of
relations as in Figure \ref{complete_samples}. We find that relations
such as the $M_{UV}$ vs. $M_H$ and the Star Formation Rate (SFR)
vs. $M_H$ do not change much across samples of different absolute
magnitude cuts. The $M_H$ vs. SHMR relations follow a common trend at
high SHMRs, but do start to change significantly as we approach the
lower end of SHMR values allowed by the magnitude cuts. This is primarily because as we go closer to the $M_{UV}$ threshold, the high mass halos are much more likely to occupy galaxies with a given SHMR compared to low mass halos, which then results in a turnover in the mean trends at low SHMR (see Figure \ref{muv_selected_samples}: Panel 3); as we increase the threshold $M_{UV}$, the turnover will occur at larger values of SHMR (or $M_H$). The effect is more apparent in the $M_H$ vs. SHMR relation compared to other relations because of 1) larger scatter (see Figure \ref{complete_samples}: Panel 3) and 2) Greater deviation from power-law in the $M_H$ vs. SHMR relation. This could
potentially affect the estimations of the $M_{H}$ vs. SHMR relation from
observational data; in the case of H16 the errorbars are large enough
that this may not be a serious issue, but it may deserve a more
careful consideration while making higher precision estimates in the
future. \par
We further note that while the trends in the mass-luminosity relations
(leftmost panel in Figure \ref{complete_samples}) in H16 are
consistent with \texttt{BLUETIDES}, the estimated halo masses at
$z=6,7$ appear to be slightly higher than what we might expect from an
extrapolation of the mean relations at $z=8,9,10$. We will discuss this
further after the next section after we move to the clustering of
the galaxy population.

\subsection{Clustering}

\begin{figure}
  \includegraphics[width=0.45\textwidth]{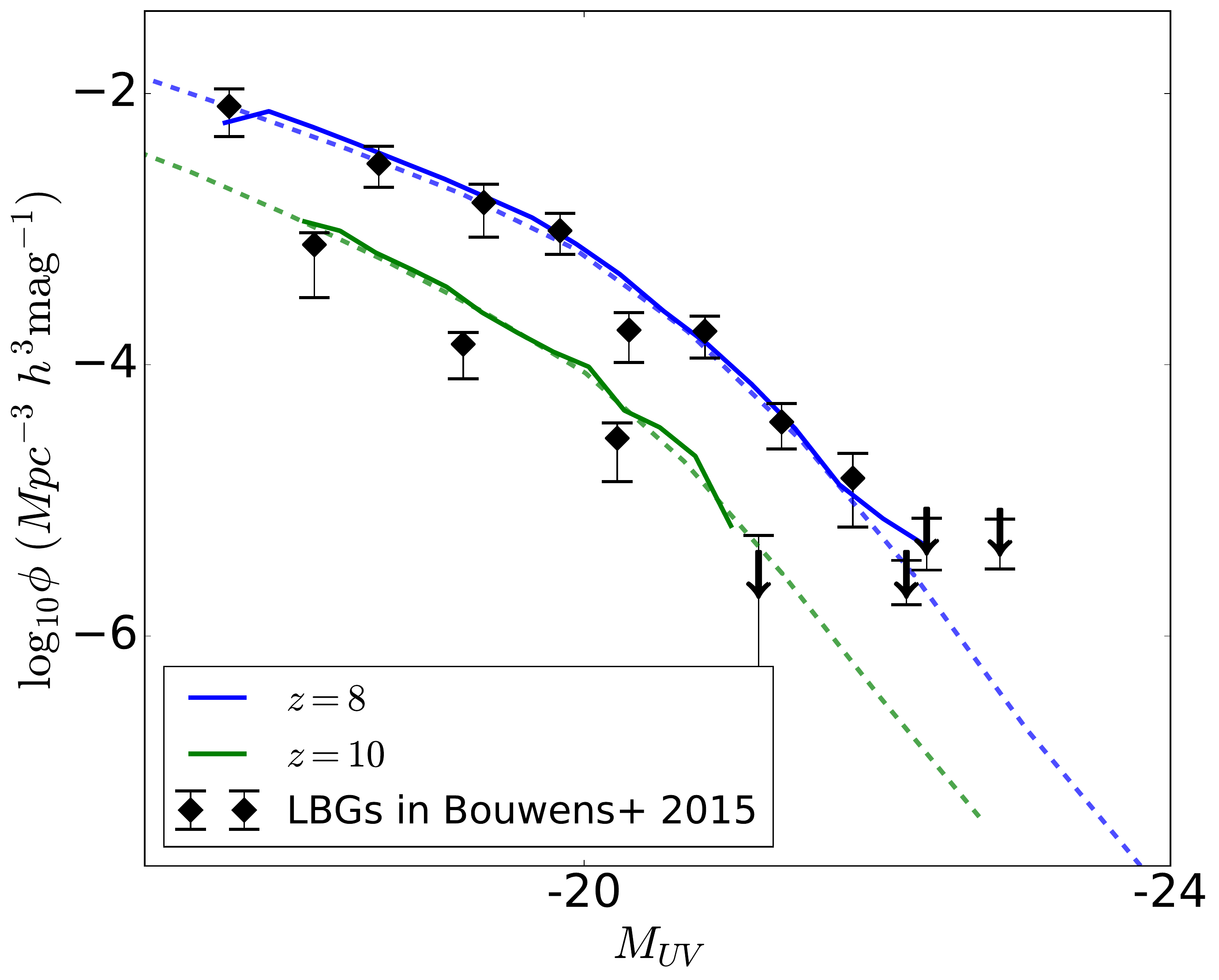}
  \caption{UV luminosity functions (LFs) at $z\sim8$ (blue) and $z\sim10$ (green) :  The black points correspond to observed UV LFs reported by B15. The dashed lines correspond to UV LFs of all galaxies in the respective target snapshot. Solid lines correspond to the UV LFs of the modulated galaxy samples extracted from the observed redshift distributions of B15 from snapshots ranging from $z=8$ to $z=13$. } 
  \label{luminosity_function}
\end{figure}


\begin{figure*}
  \includegraphics[width=18cm,height=8.5cm]{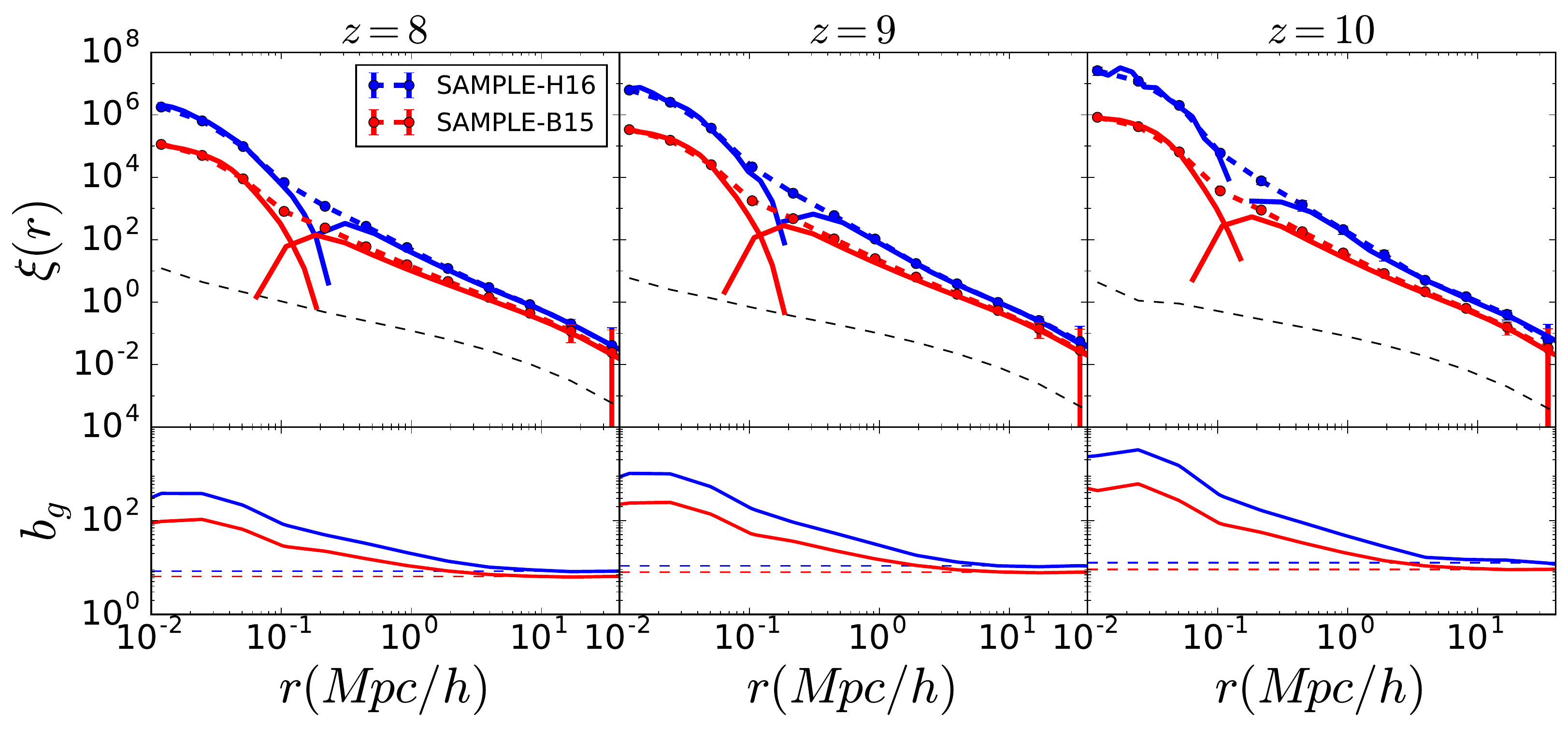}
  \caption{Spatial clustering of galaxies predicted by
    \texttt{BLUETIDES} at redshifts 8,9 and 10 from left to right for
    galaxy samples of different limiting magnitides $m_{UV}$ corresponding to \protect\cite{2015ApJ...803...34B} (hereafter B15) and \protect\cite{2016ApJ...821..123H} (hereafter H16)  in red and blue lines respectively. TOP PANELS: Solid lines represent the one
    halo and two halo contributions of the two point galaxy
    correlation function. The colored dashed lines are the total correlation functions. Thin black dashed lines are the dark matter correlation functions. BOTTOM PANELS: Scale dependent bias
    at redshifts 8,9,10. The dashed lines are the large scale
    linear biases.}
\label{spatial_clustering_H16}
\end{figure*}

In this section, we investigate the clustering predictions of \texttt{BLUETIDES}. We select galaxies
in \texttt{BLUETIDES} for galaxy samples with different limiting apparent magnitudes pertinent to the current detection limits. We consider a flux limit from the deepest field (XDF) compiled in B15 ($H_{160}\lesssim29.5$), for which B15 investigated the UV luminosity functions. This flux limit roughly corresponds the faint end in the observed UV luminsoity functions in B15. We shall refer to this sample as SAMPLE-B15. Additionally, note that \cite{2016ApJ...821..123H} (hereafter H16) extracted two considerably bright galaxy subsamples ($M_{UV}\lesssim-19.3,-19.5$) and performed clustering measurements at $z\sim7$; \cite{2014ApJ...793...17B} hereafter B14, does the same with similarly bright galaxy samples. While we are yet to have an observed clustering measurement at $z\sim8$, we do seek to investigate whether the trends in the redshift evolution are consistent with current frontiers of clustering measurements. We also therefore additionally consider the limiting magnitudes consistent with one of the bright samples of H16 (flux cuts corresponding to absolute UV magnitude cuts, $M_{UV}<-19.3$ at $z\sim7$) and refer to it as SAMPLE-H16. As discussed in section \ref{simulation}, Figure \ref{screenshot} illustrates how galaxies within current detection limits are distributed in a particular slice w.r.t. the dark matter density field. 

B15 selects galaxies at different photometric redshift bins to select LBGs using color selection criteria. In order to ensure that our galaxies selected at target snapshots $z=8, 10$ represent LBGs, we look at the UV luminosity functions at redshifts 8 and 10 in Figure \ref{luminosity_function}. The UV luminosity functions for all galaxies in the respective target snapshots (dashed lines) are in reasonable agreement with the observed UV luminosity functions of B15 as also shown in \cite{2016MNRAS.455.2778F}. We modulate our redshift distribution to represent the B15 distributions at $z\sim8$ and $z\sim10$ by subsampling galaxies across snapshots $z\sim8$ to 13, in accordance to the observed distributions. The solid lines show the UV Luminsoity functions for the modulated sample of galaxies. Very importantly, we find that the resulting UV Luminosity functions for our modulated galaxy samples are consistent with the ones obtained from galaxies in a single target snapshot. This implies that the spatial correlation functions can be obtained using all candidate galaxies at the respective target snapshots, and the modulated redshift distribution can be used for the angular correlation functions while performing the Limber projection using Eq. \ref{limber_from_xi}.

\subsubsection{Spatial Clustering of galaxies in \texttt{BLUETIDES}}
Figure \ref{spatial_clustering_H16} (top panels)
shows the spatial correlation function of galaxies in
\texttt{BLUETIDES} (dashed lines) with the one halo and two halo
contributions shown separately (solid lines) as well as the dark
matter at $z=8,9,10$. The bottom panels show the associated scale
dependent bias, $b_g$ (defined as the square root of the ratio of the
galaxy over dark matter correlation). The errors bars are delete-1
jackknife errors with each jack-knife subsample corresponding to a
subvolume obtained by removing one of the octants from the full volume
of \texttt{BLUETIDES}. The one-halo term dominates at scales $r< 0.12$
Mpc/$h$ above which the two halo term takes over. As expected, the brighter SAMPLE-H16 galaxies are more strongly clustered; furthermore, also note that clustering is steeper for the more luminous SAMPLE-H16 galaxies; therefore, as we go to smaller scales the clustering becomes more and more sensitive to the flux limit, with the one-halo clustering showing a significantly steeper dependence on the limiting magnitude as compared to two-halo clustering. These features have been found in observational studies of clustering dependence on luminosity \citep{2006ApJ...637..631K,2011ApJ...737...92S,2016ApJ...821..123H}.

A significant scale dependence in the galaxy bias is seen at scales
$0.5\lesssim r \lesssim10$ Mpc/$h$ (well within the two halo regime) indicating the presence of non-linear effects at
scales larger than the length scales associated with individual host
halos. Evidence of scale dependence in the halo bias has been previously reported using N-body simulations \citep{2009MNRAS.394..624R,2016MNRAS.463..270J}. \cite{2016MNRAS.463..270J} (hereafter J16) uses simulations with volumes comparable or larger than \texttt{BLUETIDES} to study non-linear bias at redshifts $z\sim2-5$. J16 reports some universal properties of (redshift independent) in the scale dependent bias and provides a fitting function. In our upcoming paper (Bhowmick et. al. 2017 in prep), we investigate whether their results agree with the predictions of \texttt{BLUETIDES} by making a precise comparison for a wide range of halo mass bins. We find that our predictions are in fairly good agreement (only slightly lower by an amount $\lesssim 0.3$ dex in general) with J16, extending the validity of J16 results to regimes well beyond $z\sim2-5$; we have discussed this in more detail in Bhowmick et. al. (2017) (in prep). The foregoing emphasize the importance of measuring clustering
directly with simulations as (we shall see next) the scales probed
by H16 mesurements fall within the scales where we are see
a significant influence on non-linear effects.
\subsubsection{Angular Clustering}

\begin{figure}
  \includegraphics[width=0.45\textwidth]{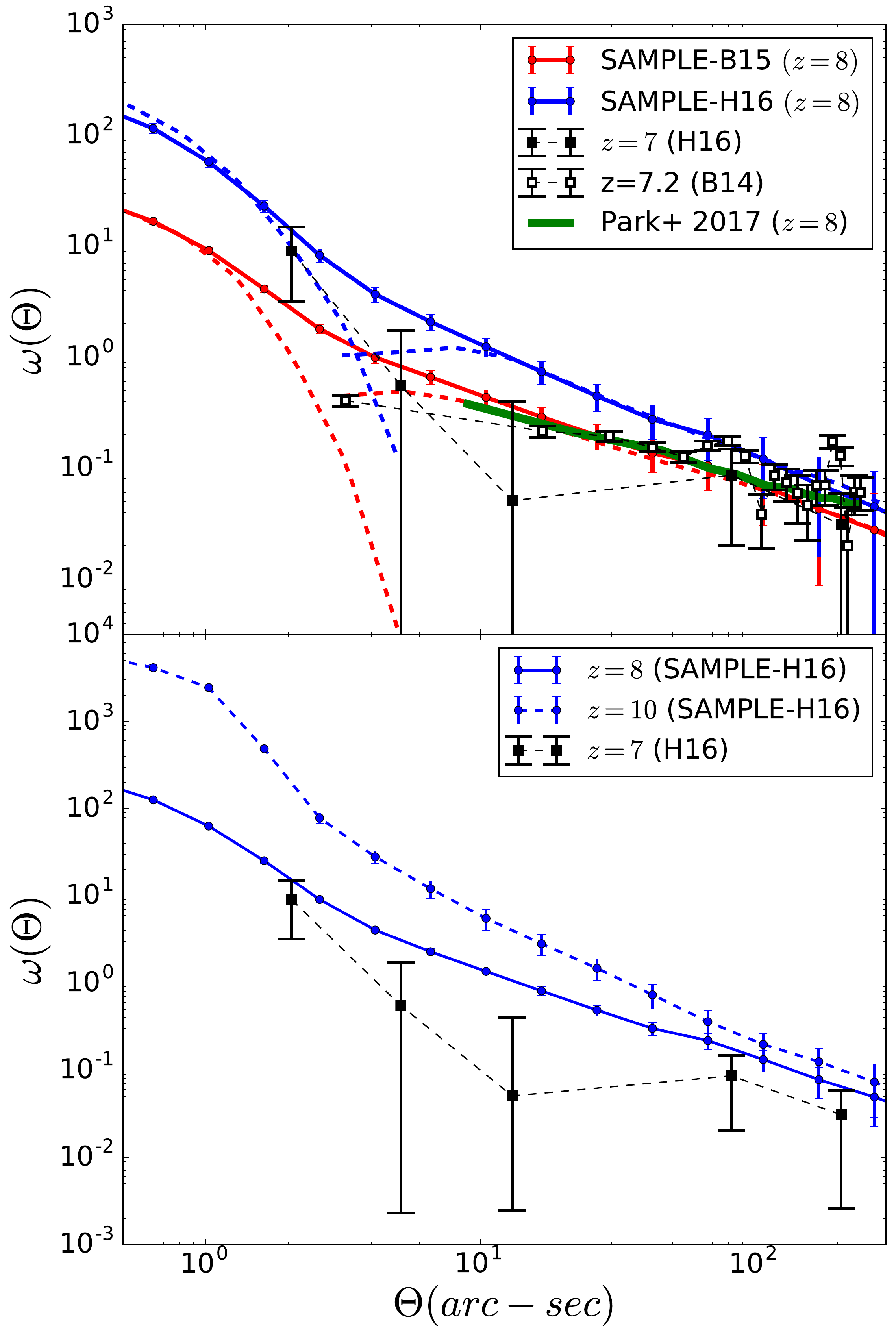}
  \caption{TOP PANEL: Angular clustering for flux limited samples at $z\sim8$: Solid line is the total correlation function whereas the dashed lines correspond to one-halo and two-halo contributions respectively. Red lines corresponds to the flux limit for SAMPLE-B15 (see Figure \ref{luminosity_function}).  The blue line corresponds to SAMPLE-H16. Green line corresponds to the measurement of \protect\cite{2017MNRAS.472.1995P}. The filled and open squares are observed measurements of H16 and  \protect\cite{2014ApJ...793...17B} (hereafter
    referred as B14) respectively at $z\sim7$, thereby identifying angular scales at which current measurements have been performed. BOTTOM PANEL: Redshift evolution of angular clustering for SAMPLE-H16 at redshifts 8-10. The blue solid and dashed lines show the angular correlation functions at $z\sim8$ and $z\sim10$ respectively. Filled squares are the measurements performed by H16 at $z\sim7$.}
  \label{angular_clustering}
\end{figure}



In this section we use Eq.~(\ref{limber_from_xi}) to project the spatial correlation functions and predict the angular clustering at these redshifts. Top panel of Figure \ref{angular_clustering} shows the predicted angular
correlation functions obtained by projecting the galaxies to $z=8$ for SAMPLE-B15 (red) and SAMPLE-H16 (blue). Note that the selection of brighter galaxies in SAMPLE-H16 can potentially steepen the redshift distribution compared to B15 distributions, we investigate this and find that the resulting uncertainty caused due to this is $\sim10-15\ \% $; this is not large enough to impact our results and we shall continue using the B15 distributions for our calculations. SAMPLE-H16 naturally has a significantly higher clustering (particularly at small scales where the difference is over an order of magnitude). Note that our results agree well with the predictions of \cite{2017MNRAS.472.1995P} at the same flux limit of SAMPLE-B15; they perform a very similar analysis using Semi-Analytical approach. They do measure a slightly higher clustering a large scales (however well within our error bars), but their simulation box is much smaller than ours ($\sim 100 Mpc/h$) thereby increasing the uncertainity at large scales. The dashed lines are the one-halo and two-halo contributions; the one-halo term dominates on scales $\Theta \lesssim 3-4$ arc-sec, and the two-halo term dominates on scales $\Theta \gtrsim 10$ arc-sec. \texttt{BLUETIDES} predictions for the angular correlations allow us
to make direct statements on the contribution of the one-halo versus
two halo term in the current measurements. In particular, we note the
B14 measurements (covering scales from $1.6$ to $1.3\times10^{2}$
arcsecs) are mostly probing the two halo regime, with only one data
point lying on the transition region between the one-halo and two-halo
regime. The H16 measurements not only probe the two-halo regime and
the transition region very well, but also contains a measurement at
scales $\sim1.3$ arcsecs which is within the one-halo regime. It is however important to note that we are yet to accurately probe the one-halo regime at these redshifts.

Bottom panel of Figure \ref{angular_clustering} shows the redshift evolution of the angular clustering and compares them to observed measurements at $z\sim7$. Overall, there is an approximately linear increase in the clustering amplitude with redshift, which becomes significantly steeper in the one-halo regime. The observational measurements of H16 are shown as solid squares which corresponds to the $z\sim7$ photometric redshift selection window ranging from $z\sim6$ to $z\sim8$.  \texttt{BLUETIDES} cannot make prediction at this range of redshifts at the moment to precisely compare with the measurements, and we reserve this for future work with upcoming BLUETIDES II. It is nevertheless noteworthy that the redshift evolution at 8 and 10 look reasonable w.r.t the observed measurement at $z\sim7$, which provides us an optimistic picture about the ability of \texttt{BLUETIDES} to describe galaxy clustering as well as other statistics \citep{2016MNRAS.455.2778F}, and consequently, the galaxy formation physics at these redshifts. Also note that \protect\cite{2017MNRAS.472.1995P} does not predict a significant change in the large scale behavior of the angular correlation from $z\sim8$ to $z\sim7$.



\subsection{Bias and typical Halo Masses}
\label{bias_and_mass}
\begin{figure*}
  \includegraphics[width=\textwidth]{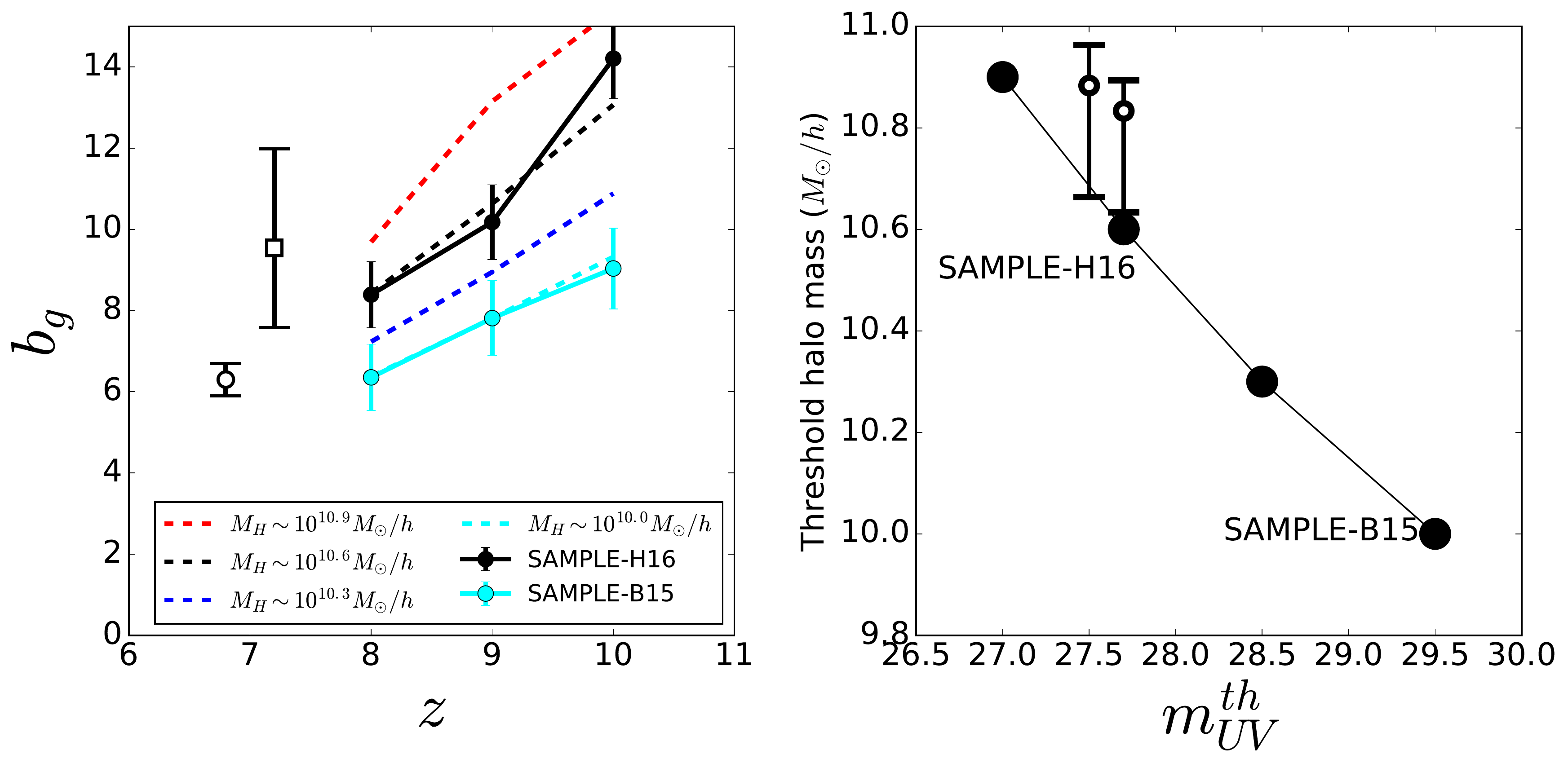}
  \caption{LEFT PANEL: Linear galaxy bias at $z=8,9,10$ as predicted by
    \texttt{BLUETIDES} for SAMPLE-B15 and SAMPLE-H16. Error bars are delete-one jack knife
    errors. Linear regime is identified to be at scales greater than
    $\sim 13$ Mpc/$h$. The dashed lines show the large scale biases for galaxy samples corresponding to various halo mass thresholds in the units of $M_{\odot}/h$. The data points at $z=7$ are measurements of
    observed galaxy samples in H16 (open cirlces: HOD
    modelling) and B14 (open squares: power law modelling). RIGHT PANEL: Estimated typical threshold halo masses at $z\sim8$ for the galaxy samples at different limiting magnitudes shown in the left panel. The open black circles are the measurements by H16 using HOD modelling at $z\sim7$.}
  \label{large_scale_bias}
\end{figure*}
Figure \ref{large_scale_bias} (left panel) shows the \texttt{BLUETIDES} predictions
(blue dots) for the large scale linear bias as a function of redshift. Note that here we show a somewhat wider range of the threshold apparent magnitudes in order to show the overall dependence on the limiting mgnitude. These are directly obtained from the galaxy
and dark matter correlation functions in Figure \ref{spatial_clustering_H16}. For SAMPLE-B15 ($H_{160}\lesssim29.5$), \texttt{BLUETIDES} predicts bias values $b_g = 5.9 \pm 0.9$, $b_g =
7.1 \pm 0.8$ and $b_g = 7.8 \pm 1.1$ at $z=8,9,10$
respectively. For SAMPLE-H16  ($H_{160}\lesssim27.7$), the bias is significantly higher with a value of $8.1\pm1.2$ at $z\sim8$. The evolution of the bias from $z=8$ to $z=10$ is $\propto (1+z)^{1.6}$ approximately; while we don't yet have a direct prediction at $z=7$,
the predicted redshift evolution seems to be consistent with the bias
value of $b_g = 6.3 \pm 0.4$ (solid black circle in Figure
\ref{large_scale_bias}) obtained by H16 at $z=7$ via HOD modelling of
their clustering measurements. H16 and B14 also estimate the bias by
fitting their clustering measurements to a power law but the agreement
of these estimates with \texttt{BLUETIDES} is worse than that obtained
using HOD modelling in H16. It is not surprising that HOD model fitting performs better than power law fitting because, as H16 correctly points out, the bias estimation using HOD modelling takes proper account of satellite galaxies by defining an effective bias $b_g^{\textrm{eff}}$ as $b_g^{\textrm{eff}}=\int d M_H N(M_H) \frac{dn}{dM_H} b_g(M_H,z)$ where $N(M_H)$ contains contributions from central as well as satellite galaxies. 

In order to estimate the typical host halo masses, we compare the large scale biases of our galaxy samples to those obtained from various halo mass thresholds. The dashed lines show the large scale biases for various halo mass thresholds $\log_{10}({M_H h/M_{\odot}})\sim 9.9,10.3,10.6,10.9$ for cyan, blue, black and red colors respectively, and how they agree with the values shown by the circles of corresponding color. The corresponding typical threshold halo mass estimates at $z\sim8$ are summarized in Figure \ref{large_scale_bias} (right panel). Thus, according to \texttt{BLUETIDES}, at $z\sim8$ the galaxies above the detection limit of B15 are hosted by halos having masses $\gtrsim10^{10} M_{\odot}/h$. Bright galaxy subsamples chosen by H16 have a threshold host halo mass $\gtrsim 4\times 10^{10} M_{\odot}/h$ at $z\sim8$. If we compare this to the halo mass threshold estimate obtained by H16 using HOD modelling, we find that our estimates lie at the lower end of the error bar. Therefore, our halo mass estimates seem to be slightly underestimated compared to them. This then relates back to Figure \ref{complete_samples}, where our mean trends also seem to be slightly lower compared to the observational data points produced by H16. In the next section we investigate as to whether \texttt{BLUETIDES} really underestimates the typical halo masses, or is it just a mere consequence of the definition of the threshold halo mass in an HOD model; to do this, we shall compare the HODs derived from \texttt{BLUETIDES} to
those adopted in H16.

\subsubsection{HODs and \texttt{BLUETIDES}}
 \label{HODs_and_bluetides}

\label{HOD_bluetides_sec}
 \begin{figure}
  \includegraphics[width=0.45\textwidth,height=11cm]{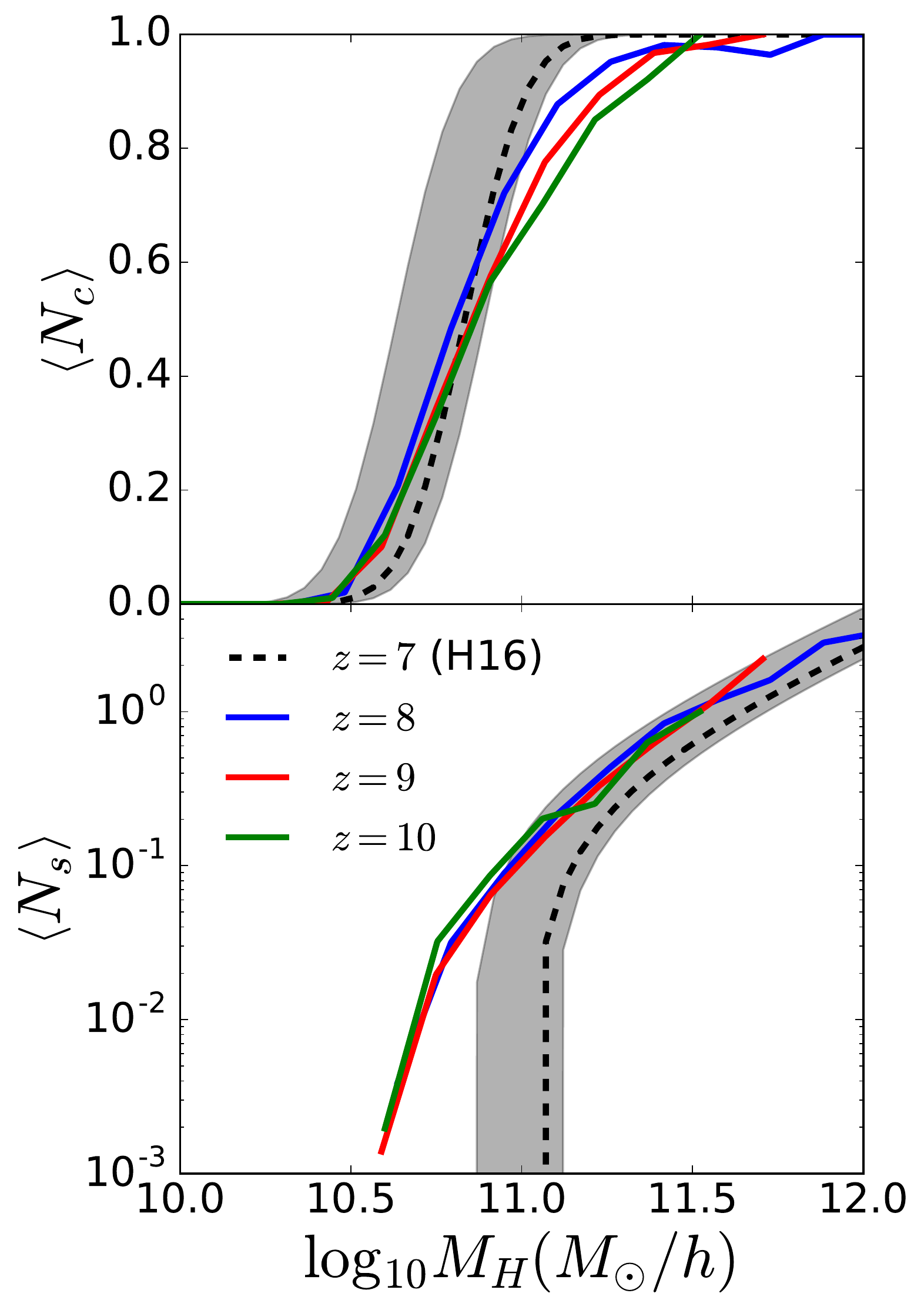}
  \caption{Halo occupation distributions for SAMPLE-H16 at $z=8,9,10$. The top and bottom panels
are mean occupations as a function of $M_H$ for central and satellite galaxies
respectively. The
dashed lines with the shaded regions are the HODs obtained by H16 using the
clustering measurements at $z=7$.}
\label{HOD_bluetides}
\end{figure}

\begin{figure}
\vbox{
    \includegraphics[width=0.45\textwidth,height=7cm]{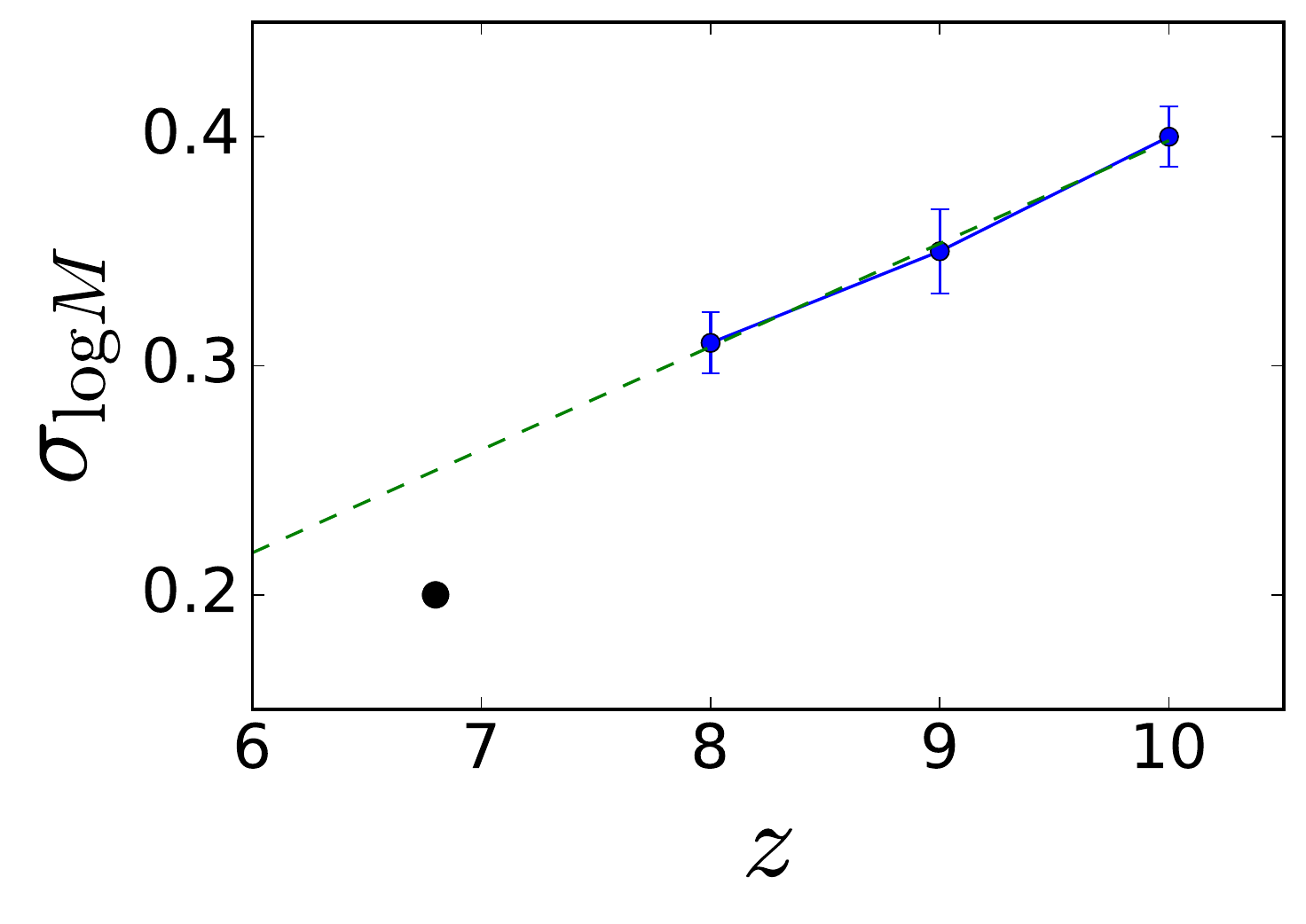}

 \caption{The variation of $\sigma_{\log M}$ with redshift.
Blue circles are the parameter values and the error bars are the square roots of
the covariance of the HOD fits. The green line is a linear fit of the data points
at $z=8,9,10$. The black circle is the value assumed by H16 for their HOD fitting to observed clustering measurements.}
\label{HOD_par1}
}

\end{figure}

To compare the results from H16 to \texttt{BLUETIDES} we now briefly
look at the HOD actually predicted from the simulations and
parametrizing it in exactly the same form as chosen by H16 (see also
Section 3).

Figure \ref{HOD_bluetides} shows the mean central and satellite occupations for SAMPLE-H16 obtained from \texttt{BLUETIDES}.  We parametrize these HODs using Eqs. (\ref{central}) and (\ref{satellite}). Note that except for the step function "width" (parametrized by $\sigma_{\log M}$) in the central galaxy occupation, the mean occupations $z=8,9,10$ do not exhibit a significant redshift evolution. Redshift evolution of $\sigma_{\log M}$ is plotted in Figure \ref{HOD_par1}. The remaining best fit HOD parameter estimates are $\log_{10}{M_{min}}\approx10.8$, $\log_{10}{M_0}\approx10.5$ and $\log_{10}{M_1}\approx11.5$ and $\alpha\approx1.2$ (all masses in units of $M_{\odot}/h$). We first note that the threshold halo mass $M_{min}$ from \texttt{BLUETIDES} has much better agreement with the estimate of H16 as we can clearly see in the top panel of Figure \ref{HOD_bluetides}; this tells us that the difference found with the threshold halo mass calculated from the bias estimates of \texttt{BLUETIDES} is merely a consequence of the definition of $M_{min}$. However, there are some other noteworthy distinctions between HODs from \texttt{BLUETIDES} and the best fit HOD obtained by H16: 
\begin{itemize}
\item \texttt{BLUETIDES} predicts a larger scatter, $\sim \sigma_{\log{M}}$=0.25 (see Figure \ref{HOD_par1}) at $z=7$ and even higher at $z\sim8,9,10$, i.e. a larger scatter for the halo mass-luminosity relation compared as opposed to $\sigma_{\log{M}}=0.2$ assumed by H16. 
\item The minimum halo mass for satellite galaxy occupancy ($M_0$) predicted by \texttt{BLUETIDES} is significantly lower compared to H16 value ($\log_{10}{M_0 h/M_{\odot}}\approx10.9$).
\item \texttt{BLUETIDES} also predicts a slightly larger $\alpha\approx 1.2$ compared to the value assumed by H16 ($\alpha=1$), implying that
higher mass halos will have higher number of satellite galaxies compared what is predicted by H16.
\end{itemize}
Note that for all of the above parameters, H16 had to assume additional relations (Eqs. 54, 55 in H16) to perform their fits at $z=7$ due to lack of statistical
accuracies; these relations are based on results based at $z=0,5$ \citep{2004ApJ...609...35K,2005ApJ...633..791Z,2006ApJ...647..201C}, and are therefore extrapolated from HODs at lower redshifts. In our upcoming paper (Bhowmick et. al. in prep), we go deeper into this aspect and perform a much more detailed investigation of the differences between HODs at low and high redshifts.

\subsection{Angular clustering: HODs vs \texttt{BLUETIDES} predictions}
\label{hod_clustering}

\begin{figure}
  \includegraphics[width=0.48\textwidth,height=12cm]{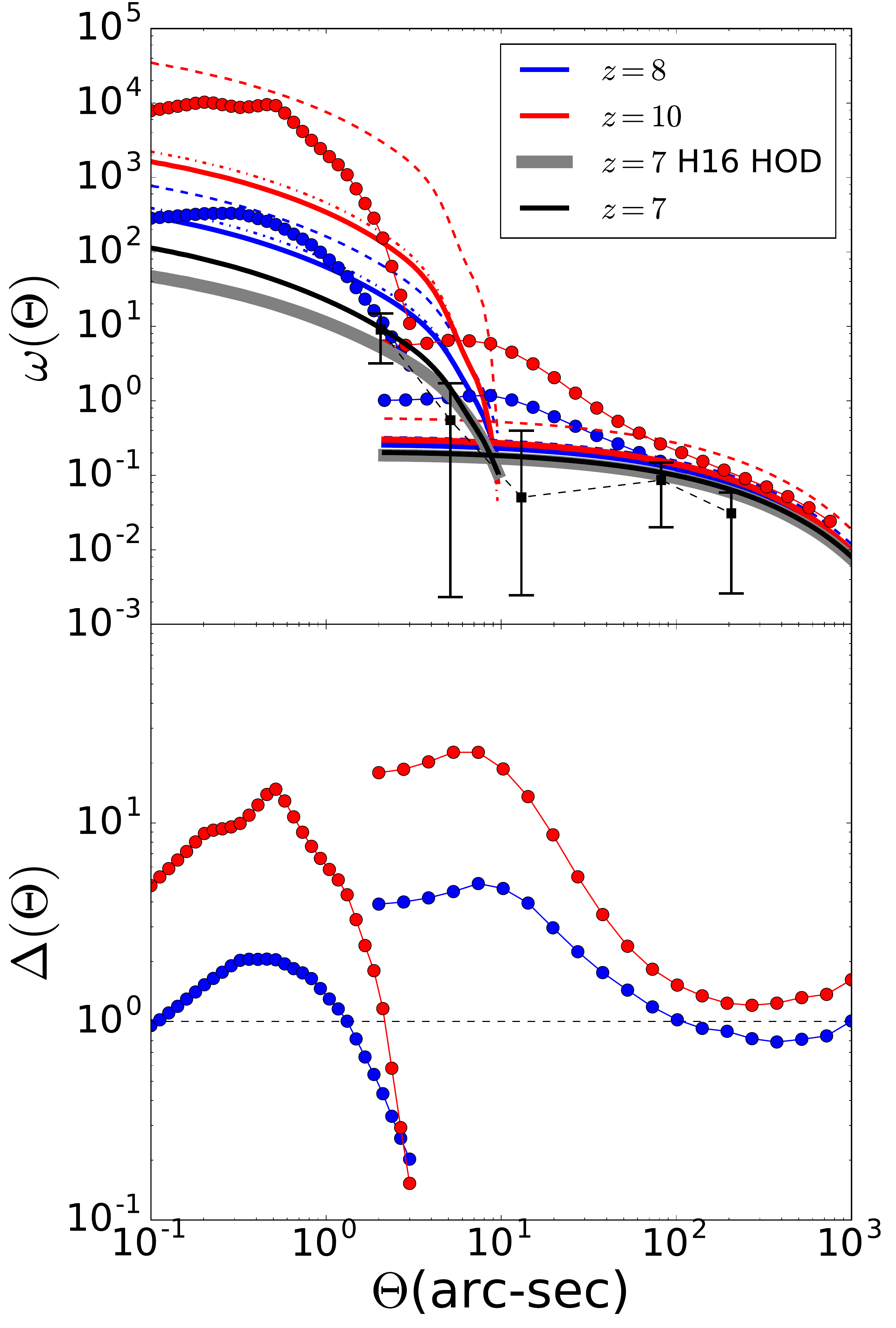}
  \caption{Comparision of the angular clustering of \texttt{SAMPLE-H16} with HOD
predictions. TOP PANEL: The filled circles represent the predictions of the one-halo and two-halo angular clustering from the simulations. The solid lines of corresponding color are the predictions of an HOD model where the HODs are shown in Figure
\ref{HOD_bluetides}. For $z=7$, the black line is the HOD model prediction
estimated via linear extrapolation of HOD parameters at $z=8-10$. The grey transparent line is the HOD prediction by the best fit HOD of H16. At $z=8,10$ there are additional dashed and dot dashed lines corresponding to $\sigma_{\log M}=0.2$ and $\alpha=1$ (low redshift values assumed by H16) respectively while keeping all the other parameters fixed at \texttt{BLUETIDES} predictions. BOTTOM PANEL: $\Delta(\Theta)$ is the ratio between the angular clustering in \texttt{BLUETIDES} vs. the angular clustering predicted by an HOD model. The dashed line denotes $\Delta(\Theta)=1$.}
\label{Harikane_HOD}
\end{figure}

We now directly compare the one-halo and the two-halo terms of the angular clustering results from \texttt{BLUETIDES} to the predictions from the reconstructed HOD
model as formulated in H16 and is described in section
\ref{HOD}. We have five HOD parameters namely $M_{\textrm{min}}$, $M_0$,
$M^{'}_{1}$, $\sigma_{\log{M}}$, $\alpha$. We
compute the one-halo and two-halo predictions of the HOD model at the
relevant redshifts.

Figure \ref{Harikane_HOD} (top panel) overlays the one-halo and
two-halo contributions of \texttt{BLUETIDES} with the HOD model
predictions at $z=7,8,10$. The solid circles represent the direct
predictions from \texttt{BLUETIDES} via pair counting. The solid lines are the HOD model predictions using HODs obtained directly by
\texttt{BLUETIDES} (HODs are shown in Figure
\ref{HOD_bluetides}). \texttt{BLUETIDES} prediction for $z=7$ HOD is
obtained via extrapolations of HOD parameters at $z=8-10$. $\sigma_{\log M}$ for $z=7$ is obtained via a linear extrapolation of the redshift evolution from $z=10$ to $z=8$. 
(shown in Figure \ref{HOD_par1}). The remaining HOD model parameters do not exhibit any significant redshift evolution as can be clearly seen in Figure \ref{HOD_bluetides}; therefore their values at $z=7$ are assumed to be the ones quoted in section \ref{HODs_and_bluetides} for $z=8-10$. The black solid line in Figure
\ref{Harikane_HOD} (top panel) shows the corresponding HOD model
clustering. Note that for $z=7$, there is an additional grey line
which represents the predicted HOD model clustering using HOD estimated from observations in H16
(dashed curves in Figure \ref{HOD_bluetides}). We find that in the two-halo regime at $z=7$, the HOD model angular clustering estimated from observations in H16, and the ones derived from \texttt{BLUETIDES}
HOD, agree. This tells us that the slight differences in the $\sigma_{logM}$ values of H16 and
\texttt{BLUETIDES} HOD at $z=7$ (discussed in section
\ref{HODs_and_bluetides}) do not have a significant effect on the two halo clustering at $z=7$; however at higher redshifts the differences may become more significant as $\sigma_{logM}$ increases. In the one-halo regime there is a noticable difference mainly because of the difference in $M_0$ in the satellite mean occupations. In addition to the solid lines for $z\sim8,10$, we also plot the dashed lines and dot dashed lines which correspond to $\sigma_{\log M}=0.2$ and $\alpha=1$ (low redshift values assumed by H16) respectively while keeping all the other parameters fixed at \texttt{BLUETIDES} predictions. $\alpha$ is only slightly different, therefore not surprisingly it does not make a significant difference (less than 5 percent) to clustering. However, $\sigma_{\log M}=0.2$ makes a substantial difference i.e. $\sim 0.4$ dex higher at $z\sim8$ and $\sim 1$ dex higher at $z\sim10$. This is expected because decreasing $\sigma_{logM}$ while keeping other parameters fixed reduces the contribution of lower mass halos, thereby increasing the clustering.

\par Interestingly, at $z=8,10$ the HOD models predict an $\omega(\Theta)$
that is different from that calculated in \texttt{BLUETIDES}. In order to quantify the differences, we define $\Delta(\Theta)\equiv\omega_{\textrm{sim}}(\Theta)/\omega_{\textrm{HOD}}(\Theta)$ where $\omega_{\textrm{sim}}(\Theta)$ is the direct prediction from simulations, and $\omega_{\textrm{HOD}}(\Theta)$ is the HOD model prediction. Figure \ref{Harikane_HOD} (botton panel) shows $\Delta(\Theta)$ at $z=8,10$.   
In the two-halo regime, HODs and simulations agree well at the largest
scales ($\Theta \sim 10^{2}$ arc-sec). As $\Theta$ decreases, while
still in the 2-halo regime, $\omega_{\textrm{sim}}(\Theta)$ is
enhanced compared to $\omega_{\textrm{HOD}}(\Theta)$ at scales of
$\Theta \sim 10$ arc-sec by a factor of upto $\Delta\sim2$ at
$z=8$. The foregoing may be due to the fact that HOD modelling does
not usually account for the scale-dependence of the bias in the
two-halo regime. However, some of the very recent analyses using HOD
models have indeed started accounting for the scale-dependence
\citep{2017arXiv170406535H}. The one-halo regime is of great interest
because non-linear effects are expected to be much more pronounced. In
the one-halo regime, $\omega_{\textrm{sim}}(\Theta)$ is enhanced
compared to $\omega_{\textrm{HOD}}(\Theta)$ by a factor of
$\Delta\sim2$ at $z=8$ at scales of $\Theta \sim 0.2$ arc-sec, while
$\omega_{\textrm{sim}}(\Theta)$ starts to drop faster than that of
$\omega_{\textrm{HOD}}(\Theta)$ at $\Theta > 1$ arc-sec. Towards the
end of the one-halo regime at $\Theta \sim 2$ arc-sec,
$\omega_{\textrm{sim}}(\Theta)$ is diminished by a factor of
$\Delta\sim2$. The foregoing differences are more pronounced at higher
redshifts. Note that in the parametrization of the HOD models the halo
profile is assumed to be NFW (see Eqs.~(\ref{eqn_1h_cs}),
(\ref{eqn_1h_ss}), (\ref{eqn_2h})) in the satellite galaxy density
profile. In practice, this implies a higher concentration of galaxies
in the inner regions of halos than expected in an NFW. This regime of
enhanced clustering has not yet been probed by current observations,
but it may have interesting implication for JWST observations (which
are likely to probe these small scales).

\section{Discussion}
\label{discussion}
In this analysis, we performed a detailed study of the clustering of high redshift
galaxies during the reionization epoch as predicted by the \texttt{BLUETIDES}
simulation. The extraordinarily high volume and high resolution of
\texttt{BLUETIDES} enabled us to study clustering from scales as small as $\sim
10^{-2}$ Mpc/$h$ to scales as large as  $\sim 10^{2}$ Mpc/$h$. We make predictions for the one-halo and two-halo clustering for flux limited galaxy samples; the magnitude limits are chosen to be detection limits of B15, as well as thresholds chosen by H16 for their clustering measurements at $z\sim7$. We find a substantially steeper dependence of clustering on redshift as well as limiting magnitude in the one-halo regime, as compared to the two-halo regime. Our redshift evolution seems to be reasonable in view of the current observational measurements at $z\sim7$. We also compare our clustering to the predictions of HOD modelling (typically used to infer bias from the observations) and discussed the possible differences
in the clustering of galaxies between direct simulation predictions and HODs.

We have shown the properties of \texttt{BLUETIDES} galaxies in the
$M_H$ vs $M_{UV}$, $M_H$ vs. SFR and $M_H$ vs. SHMR planes, and found
that the H16 galaxies are well represented sub-sample in the
simulations.  (for H16 the halo masses are inferred from the
clustering via HOD model fits).  The redshift evolution of relations
such as $M_H$ vs.  $M_{UV}$, $M_H$ vs. SFR and $M_H$ vs. SHMR at
$z=6,7$ agree with the extrapolated trends predicted by
\texttt{BLUETIDES} at $z=8,9,10$. Notably, while there is an increase
in the SHMR evolution with redshift found by H16 at $4<z<7$ (which
they claim to be inconsistent with the hydrodynamic simulation in
\cite{2014ApJ...780..145T} and the semi analytical model in
\cite{2015MNRAS.453.4337S}, our simulation predicts a continued
increase in SHMR for $z=8,9,10$ in agreement with H16. As discussed by
H16 in detail, this indicates that the stellar mass assembly at $z>4$
is not efficient enough to keep up with the increase in the dark
matter mass due to dark matter accretion and halo mergers. For a
detailed discussion of the possible mechanisms of SHMR evolution, we
refer the reader to section 7.1 in H16.\par The clustering measured
from \texttt{BLUETIDES} are in good agreement with the angular
correlation in H16 and B14.  Comparing \texttt{BLUETIDES} predictions
with these observational measurements reveal that at these redshifts,
the Lyman Break Galaxy (LBG) catalogs are large enough to probe the
two-halo regime and the transition between the one-halo and two-halo
regime. H16 measurements have started to probe the one-halo clustering
but larger galaxy catalogs shall be needed to go deeper into this
regime. \par

Using the simulations we have shown the presence of significant non-linear effects
(scale dependence of the bias) at non-linear scales ($r\lesssim0.5$ Mpc/$h$) in the one-halo regime, as well as the "quasi-linear" scales ($0.5\lesssim r \lesssim 10$ Mpc/$h$) in the two-halo regime ($r>0.2$ Mpc/$h$), as have already been shown in  \cite{2016MNRAS.463..270J}. These non-linear effects are very often not taken
into account in analytical models for the two-halo correlation,
although very recent studies have started incorporating the scale
dependence in the two-halo regime \citep{2017arXiv170406535H,2017MNRAS.469.4428J} into clustering analysis of observed galaxies and found that accounting for non-linear halo bias can significantly affect the estimates of galaxy bias and halo masses. One must note that most of the previous works are based on N-body simulations, whereas \texttt{BLUETIDES} is a high resolution hydrodynamic simulation (includes baryons) which can \textit{directly} probe the non-linear clustering properties of galaxies; in our next paper (Bhowmick et. al. 2017 in prep.) we further examine the behavior of non-linear galaxy bias. \par

By fitting the galaxies in \texttt{BLUETIDES} to same parametrization
of the HOD chosen by H16 we have shown that the HODs predicted by
\texttt{BLUETIDES} have reasonable agreement with the best fit HOD
obtained by H16 within their error bars. But notably, due to higher $\sigma_{\log M}$, there seems to
be a slightly increased contribution of lower mass halos in the galaxy
sample as compared to what is assumed by H16; ignoring these low mass halos can increase the clustering amplitude by upto $\sim 0.4$ dex at small scales at $z\sim8$, this effect continues to increase at higher redshifts due to even higher  $\sigma_{\log M}$. Incidentally,
\texttt{BLUETIDES} also predicts that the higher mass halos occupy slightly 
larger number of satellite galaxies (slightly higher $\alpha$) compared to what
is assumed in H16 (this feature has also been reported at $z\sim1$ in \cite{2014MNRAS.438..825K} using semi-analytical modelling, who also point out that it may be attributed to possible uncertainities in the galaxy formation models). For the samples considered in this work, the difference in $\alpha$ is very small and does not significantly affect the clustering amplitude; we shall show in our next paper (Bhowmick et. al. in prep) that $\alpha$ continues to increase for even brighter subsamples.

At $z=8,9,10$, interestingly we find that even if we use the HODs
predicted by \texttt{BLUETIDES}, HOD modelling does not
self-consistently reproduce the shape or the redshift evolution of the
one-halo term, while the agreement in the two-halo regime is
better. This could be related to some of the intrinsic assumptions
often made in HOD modelling:

\begin{itemize}
\item A commonly made assumption is that the density profile of
  satellite galaxies trace the NFW profile of the underlying dark
  matter distribution. While this may be true in general, it may not
  be so for galaxies above certain luminosity thresholds.  This may be
  one of the reasons for why the shapes of the HOD one halo terms are
  different compared to the ones obtained by direct pair counting.

\item Even if the NFW profile is correct, the mass concentration relations which are
often extrapolated from low redshift data, might not be valid at higher redshifts.

\item Yet another commonly used assumption is that the
  satellite galaxy distribution is Poisson. Deviations from Poisson
  distribution can affect the best fit estimates of the HOD parameters
  \citep{2000MNRAS.311..793B,2000MNRAS.318.1144P,2000MNRAS.318..203S,2002ApJ...575..587B}.

\end{itemize} All of the above points will be subjected to detailed
investigations in future work where we plan to build full HOD models
(testing all of the assumptions above) using the simulations
directly. Upcoming WFIRST and JWST observations are going to provide
very large samples of galaxies in these high redshift regime. Thus, in
the light of these upcoming observations, as samples become larger and
measurements become more accurate, it will be crucial for us to
carefully understand the clustering of high redshift galaxies using detailed
simulations.

\section{Conclusion}

We predict clustering of high redshift galaxies using the \texttt{BLUETIDES} simulation for galaxy samples limited by apparent magnitude. The bias of faint galaxy samples detected in the
  Hubble legacy field ($H_{160} \lesssim29.5$) is $5.9\pm0.9$ corresponding to halo
  masses $M_H \gtrsim 10^{10} M_{\odot}$. Galaxies with
  magnitude $M_{UV} < -19.3$ (in the brighter galaxy subsamples with
  observed clustering measurements at $z \sim 6.8$) have a bias of $8.1 \pm 1.2$ at $z=8$,
  and imply halo masses $M_H\gtrsim 6\times 10^{10}$
  $M_\odot$, which is consistent with the estimates made using HOD modelling of observed samples. Thus,
despite the lack of adequate statistics in the high redshift regime,
HOD modelling performs well in estimating the host halo
masses. However, \texttt{BLUETIDES} clustering also contains features
which cannot be captured by HOD modelling. Due to non-linear halo bias, in the two-halo regime we report enhanced enhanced galaxy clustering at $\Theta \sim 10.0$ arc-sec compared to linear HOD-model. The non-linear bias predicted by \texttt{BLUETIDES} is in excellent agreement with analytical fits provided by previous works based on N-body simulations, and could therefore be easily incorporated in an HOD model. However, discrepancies between simulations and HOD models are also found in the one-halo regime where galaxy clustering can be determined independent of halo clustering. We report enhanced
galaxy clustering at $\Theta \sim 0.2$ arc-sec, and suppressed clustering at $\Theta \sim 3.0$ arc-sec compared to standard HOD model; this warrants rigorous testing of the standard HOD model assumptions before they are applied to high redshift galaxies. The foregoing discrepancies tend to increase at higher redshifts. These features have
not yet been captured by current high redshift clustering
measurements, but will be so by upcoming satellites such as the WFIRST
and the JWST.

\section{Acknowledgements}
We express our gratitude to Peter Behroozi for providing
\texttt{ROCKSTAR-GALAXIES}, and Yao-Yuan Mao for helping us set up of
the code. We thank Ananth Tenneti for discussions and help with
visualization. Special thanks to Yuichi Harikane for providing useful
insights on HOD modelling. We acknowledge funding from NSF
ACI-1614853, NSF AST-1517593, NSF AST-1616168 and the BlueWaters PAID
program. The \texttt{BLUETIDES} simulation was run on the BlueWaters
facility at the National Center for Supercomputing Applications.

\bibliographystyle{mnras}
\bibliography{references}

\end{document}